\documentclass[aps,amsfonts,pra,twocolumn,showpacs]{revtex4}
\usepackage{epsfig,amsmath,amssymb,bm,epsf,graphicx,psfrag,float}
\usepackage[all]{xy}
\usepackage{color}

\def\bra#1{\langle#1\vert}
\def\ket#1{\vert#1\rangle}
\def\ketbra#1{\vert#1\rangle\langle#1\vert}

\def\Longarrow{\protect\@lra}
\def\@lra{\relbar\joinrel\relbar\joinrel\relbar\joinrel%
          \relbar\joinrel\rightarrow}

\newcommand{\bc}{\begin{center}}
\newcommand{\ec}{\end{center}}
\newcommand{\be}{\begin{equation}}
\newcommand{\ee}{\end{equation}}
\newcommand{\bea}{\begin{eqnarray}}
\newcommand{\eea}{\end{eqnarray}}

\newcommand{\ncd}{\newcommand}
\ncd{\QCcns}{$QC_{\cal{C}}$}
\ncd{\QCc}{$QC_{\cal{C}}\;$}

\definecolor{libl}{cmyk}{0.2,0.1,0,0}

\begin{document}

\title{ Affleck-Kennedy-Lieb-Tasaki state on a honeycomb lattice is a
universal quantum computational resource}

\author{Tzu-Chieh Wei}
\affiliation{Department of Physics and Astronomy, University of British
Columbia, Vancouver, British Columbia V6T 1Z1, Canada}
\author{Ian Affleck}
\affiliation{Department of Physics and Astronomy, University of British
Columbia, Vancouver, British Columbia V6T 1Z1, Canada}
\author{Robert Raussendorf}
\affiliation{Department of Physics and Astronomy, University of British
Columbia, Vancouver, British Columbia V6T 1Z1, Canada}
\date{\today}

\begin{abstract}
Universal quantum computation can be achieved by simply performing
single-qubit measurements on a highly entangled resource state, such as
cluster states. The family of Affleck-Kennedy-Lieb-Tasaki (AKLT) states has
recently been intensively explored and shown to provide restricted
computation. Here, we show that the two-dimensional AKLT state on a honeycomb
lattice is a universal resource for measurement-based quantum computation.
\end{abstract}
 \maketitle

 {\it Introduction\/}. Quantum computation promises exponential
speedup over classical computation by exploiting the quantum
mechanical nature of physical processes~\cite{NielsenChuang00}. In
addition to the standard circuit model based on unitary evolution,
surprisingly, local measurement alone provides the same power of
computation, given only a prior sufficiently entangled
state~\cite{Oneway,Oneway2}. For this model of measurement-based
quantum computation (MBQC), universal resource states are known to
be very rare~\cite{Gross1}, but examples do exist~\cite{Cluster,
Gross, Verstraete,Cai}. The 2D cluster state on the square lattice
is a universal resource state~\cite{Oneway,Cluster}. Cluster states
can be created by the Ising interaction from unentangled states
\cite{Cluster, coldatom}, but they do not arise as unique ground
states of two-body interacting Hamiltonians~\cite{Nielsen}. However,
by careful design of Hamiltonians, certain ground states can be used
for universal MBQC~\cite{Chen,Cai10}, and this opens up an appealing
possibility of creating universal resource states by cooling.

A new perspective on MBQC emerged when it was discovered that the
one-dimensional Affleck-Kennedy-Lieb-Tasaki (AKLT)
state~\cite{AKLT}, originally constructed in the setting of
condensed matter physics, can serve as resources for restricted
computations~\cite{Gross,Brennen,Chen10}.  In any dimension, the
AKLT state is the ground state of a particularly simple Hamiltonian
which only has nearest-neighbor two-body interactions, is
rotationally invariant in spin space and shares all spatial
symmetries of the underlying lattice~\cite{commentgap}. The
discovery of the resourcefulness of AKLT states creates additional
avenues for its experimental realization~\cite{Resch}, and has
instilled novel concepts in MBQC, such as the renormalization group
and the holographic principle~\cite{Bartlett,Miyake}. However, to
date one crucial element was missing in this direction: the AKLT
family was not known to contain a {\em{universal}} resource. Here,
we overcome this gap by demonstrating that the AKLT state on a
two-dimensional honeycomb lattice is a universal resource for
measurement-based quantum computation.

 To do this, we proceed in three steps. First, we
show that it can be mapped to a random planar graph state
$|G({\cal{A}})\rangle$ by local generalized measurement, with the
graph $G({\cal{A}})$ depending on the set ${\cal A}$ of measurement
outcomes on all sites (defined below).  Second, we argue that the
computational universality of a typical resulting graph state
$|G({\cal{A}})\rangle$ hinges solely on the connectivity of
$G({\cal{A}})$, and is thus a percolation problem.  Third, we
demonstrate via Monte Carlo simulation that the typical graphs
$G({\cal{A}})$ are indeed deep in the supercritical phase.

\begin{figure}
  \begin{tabular}{ll}
    (a) & (b)\\
    \hspace*{0.1cm} \includegraphics[height=3cm]{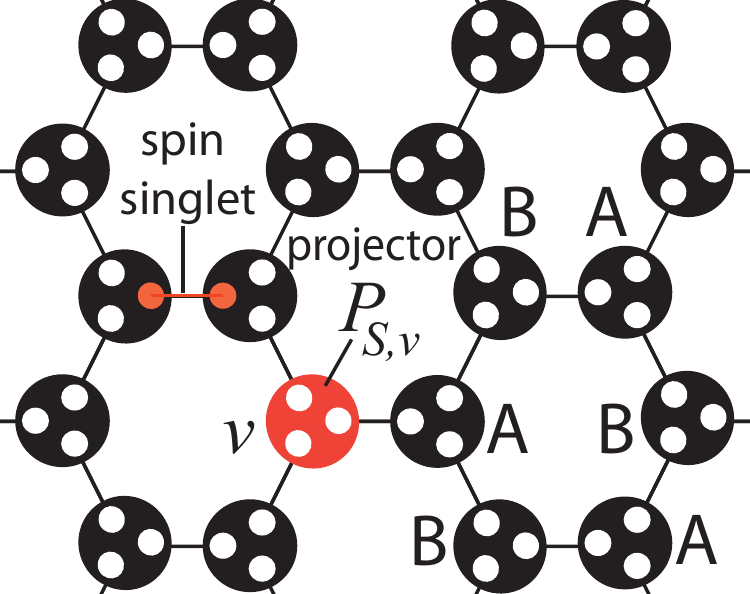} & \includegraphics[height=3cm]{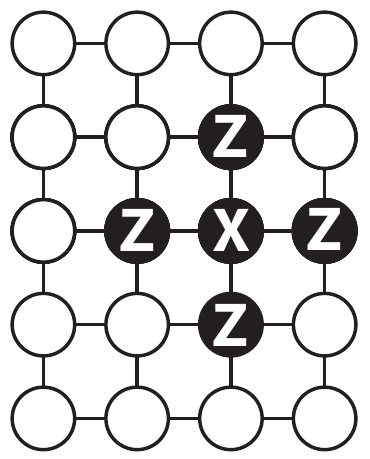}
\end{tabular}
  \caption{\label{fig:states}
  Illustrations of the AKLT state  and the 2D cluster state. (a) AKLT state. Spin singlets of two virtual spins 1/2 (i.e. qubits) are located
  on the edges of the honeycomb lattice. A projection $P_{S,v}$ at each lattice site $v$
  onto the symmetric subspace of three virtual spins defines the AKLT
  state. In one hexagon, sites are labeled by $A$ or $B$ to show the
  bi-partite (or bi-colorable) property of the honeycomb lattice.
  (b) 2D Cluster state. One qubit  resides at each lattice site and one stabilizer generator is shown.}
\end{figure}
 The AKLT state~\cite{AKLT} on the honeycomb lattice ${\cal{L}}$ has
one spin-3/2 per site of ${\cal{L}}$. The state space of each spin 3/2 can be
viewed as the symmetric subspace of three virtual spin-1/2's, i.e., qubits. In
terms of these virtual qubits, the AKLT state on ${\cal{L}}$ is (see
Fig.~\ref{fig:states}a)
\begin{equation}
  \label{AKLT2}
  |\Phi_{\rm AKLT}\rangle\equiv\bigotimes_{v \in V({\cal{L}})}P_{S,v}
\bigotimes_{e \in E({\cal{L}})} |\phi\rangle_e,
\end{equation}
where $V({\cal L})$ and $E({\cal L})$ to denote the set of vertices
and edges of ${\cal L}$, respectively. $P_{S,v}$ is the projection
onto the symmetric (equivalently, spin 3/2) subspace at site $v$ of
${\cal{L}}$~\cite{supp}. For an edge $e=(v,w)$, $|\phi\rangle_{e}$
denotes a singlet state, with one spin 1/2 at vertex $v$ and the
other at $w$.

 A graph
state $\ket{G}$ is a stabilizer state~\cite{Hein} with one qubit per
vertex of the graph $G$ and  is the unique eigenstate of a set of
commuting operators~\cite{Cluster}, usually called the stabilizer
generators~\cite{Stabilizer},
\begin{equation}
\label{eqn:stabilizerGen} X_v\bigotimes_{u\in {\rm nb}(v)}
Z_u\,\, \ket{{G}}=\ket{{G}}, \ \forall v\in V({G}),
\end{equation}
where ${\rm nb}(v)$ denotes the neighbors of vertex $v$, and
$X\equiv \sigma_x$, $Y\equiv\sigma_y$ and $Z\equiv\sigma_z$ are the
 Pauli matrices. A cluster state is a special case of graph states,
with the underlying graph being a regular lattice (see
Fig.~\ref{fig:states}b). Any 2D cluster state is a universal
resource for measurement-based quantum
computation~\cite{Oneway,Universal}.

\begin{table}
  \begin{tabular}{c|r|r|r}
    {POVM outcome\vspace{0mm}} & \multicolumn{1}{c|}{$z$} & \multicolumn{1}{c|}{$x$} & \multicolumn{1}{c}{$y$} \\ \hline
    {stabilizer generator} & $\lambda_{i}\lambda_{j} Z_iZ_{j}$, & $\lambda_{i}\lambda_{j} X_iX_{j}$ & $\lambda_{i}\lambda_{j}Y_iY_{j}$ \\
    $\overline{X}$ & $\bigotimes_{j = 1}^{3|{\cal{C}}|} X_j$ & $\bigotimes_{j = 1}^{3|{\cal{C}}|} Z_j$ & $\bigotimes_{j = 1}^{3|{\cal{C}}|} Z_j$ \\
    $\overline{Z}$ & $\lambda_i Z_i$ & $\lambda_i X_i$ & $\lambda_i Y_i$
  \end{tabular}
 \caption{\label{tbl:coding2} The dependence of stabilizers and encodings
  on the local POVM outcome. $|{\cal C}|$ denotes the total number of sites contained in a
  domain and $i,j = 1\, ..\, 3|{\cal{C}}|$. The honeycomb lattice ${\cal{L}}$ is  bi-partite and all sites can be
  divided into either $A$ or $B$ sublattice, $V({\cal{L}}) = A \cup B$; see Fig.~\ref{fig:states}a.
  One choice of the sign is $\lambda_i = 1$ if the virtual qubit $i \in v \in A$ and  $\lambda_i = -1$ if
   $i \in v' \in B$; this is due to the negative sign in the stabilizer generator for a singlet, i.e.,
   $(-\sigma_{\mu,i}\sigma_{\mu,j})|\phi\rangle_{ij}= |\phi\rangle_{ij}$ for an edge $(i,j)$.}
\end{table}

 {\it Reduction to a graph state\/}. To show that the 2D AKLT state
of four-level spin-3/2 particles can be converted to a graph state
of two-level qubits, we need to preserve a local two-dimensional
structure at each site. This is achieved by a local generalized
measurement~\cite{NielsenChuang00}, also called
positive-operator-value measure (POVM), on every site $v$ on
${\cal{L}}$. The POVM consists of three rank-two elements
\begin{subequations}
\label{POVM2}
  \begin{eqnarray}
\!\!\!\!\!\!\!\!\!\!{F}_{v,z}&=&\sqrt{\frac{2}{3}}(\ketbra{000}+\ketbra{111}) \\
\!\!\!\!\!\!\!\!\!\!{F}_{v,x}&=&\sqrt{\frac{2}{3}}(\ketbra{+++}+\ketbra{---})\\
\!\!\!\!\!\!\!\!\!\!{F}_{v,y}&=&\sqrt{\frac{2}{3}}(\ketbra{i,i,i}+\ketbra{-\!i,-\!i,-\!i}),
\end{eqnarray}
\end{subequations}
where $|0/1\rangle$, $\ket{\pm}\equiv(\ket{0}\pm\ket{1})/\sqrt{2}$
and $\ket{\pm i}\equiv (\ket{0}\pm i\ket{1})/\sqrt{2}$ are
eigenstates of $Z$, $X$ and $Y$, respectively, and $|0\rangle \equiv
|\uparrow\rangle$, $|1\rangle \equiv |\downarrow\rangle$.
Physically, $F_{v,a}$ is proportional to a projector onto the
two-dimensional subspace spanned by the $S_a=\pm 3/2$ states. The
above POVM elements obey the relation $\sum_{\nu \in
\{x,y,z\}}F^\dagger_{v,\nu} F_{v,\nu} = P_{S,v}$, i.e., project onto
the symmetric subspace, as required. The outcome $a_v$ of the POVM
at site $v$ is random, which can be $x$, $y$ or $z$, and it is
correlated with the outcomes at other sites due to the entanglement
in the AKLT state~\cite{AKLT,Katsura}. As demonstrated below, the
resulting quantum state, dependent on the random POVM outcomes
${\cal{A}}=\{a_v, v\in V({\cal{L}})\}$,
\begin{equation}
  \label{CA}
  |\Psi({\cal{A}})\rangle = \bigotimes_{v \in V({\cal{L}})} \!\! F_{v,a_v}\, |\Phi_{\rm AKLT}
\rangle= \bigotimes_{v \in V({\cal{L}})} \!\! F_{v,a_v} \bigotimes_{e \in
E({\cal{L}})} |\phi\rangle_e
\end{equation}
is equivalent under local unitary transformations to an encoded
graph state $\overline{|G({\cal{A}})\rangle}$.  We show that the
corresponding graph $G({\cal{A}})$ is constructed from graph ${\cal
L}$ by applying the following two rules:
\begin{itemize}
\item[R1]{(Edge contraction): Contract all edges $e \in E({\cal{L}})$ that connect sites with the same POVM outcome.}
\item[R2]{(Mod-2 edge deletion): In the resultant multi-graph, delete
all edges of even multiplicity and convert all edges of odd
multiplicity into conventional edges of multiplicity 1.}
\end{itemize}
A set of sites in ${\cal{L}}$ that are contracted into a single
vertex of $G({\cal{A}})$ by rule R1 is called a {\em{domain}}. Each
domain supports a single encoded qubit. The stabilizer generators
and the encoded operators for the resulting codes are summarized in
Table~\ref{tbl:coding2}.

\begin{figure}
  \hspace*{0.1cm}
  \includegraphics[width=5.5cm]{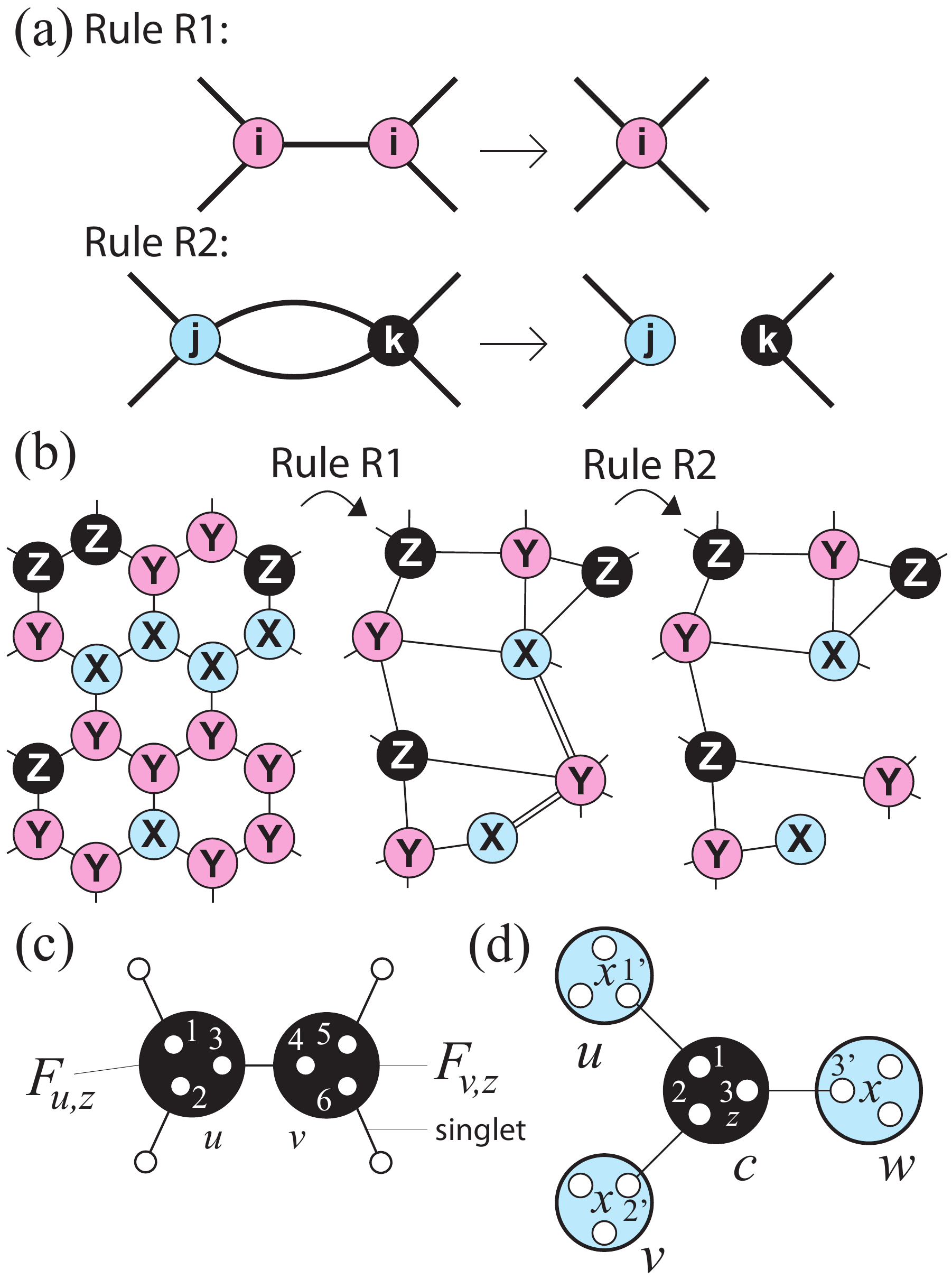}\vspace{-0.5cm}
  \caption{\label{merge}Graphical rules for transformation of the lattice ${\cal{L}}$
  into the graph $G({\cal{A}})$, depending on the POVM outcomes ${\cal{A}}$.
  a) Rules R1 and R2.
  b) An example of obtaining $G({\cal{A}})$. The alphabets inside the circle indicate the POVM outcomes.
  c) An example to illustrate the qubit encoding of a domain. d) An example for demonstrating the
  stabilizer generator.}
\end{figure}

Rule R1 derives intuitively from the antiferromagnetic property of
the AKLT state: neighboring spin-3/2 particles must not have the
same $S_a=3/2$ (or -3/2) configuration~\cite{AKLT}. Hence, after the
projection onto $S_a=\pm3/2$ subspace by the POVM, the
configurations for all sites inside a domain can only be
$\ket{3/2,-3/2,\dots}$ or $\ket{-3/2,3/2,\dots}$, and these form the
basis of a single qubit. This can also be understood in terms of the
stabilizer.
 Consider the case where two neighboring POVMs yield the same
outcome, say $z$; see Fig.~\ref{merge}c. Due to the projections
$F_{u,z}$ and $F_{v,z}$ (with $u=\{1,2,3\}$ and $v=\{4,5,6\}$ each
containing three virtual qubits), the operators $Z_1Z_2$, $Z_2Z_3$,
and $Z_4Z_5$, $Z_5 Z_6$ become stabilizer generators of
$|\Psi({\cal{A}})\rangle$. Moreover, the stabilizer $-Z_3Z_4$ of
$|\phi\rangle_{34}$ commutes with
 $F_{u,z}\otimes F_{v,z}$, and thus remains a stabilizer element
for $|\Psi({\cal{A}})\rangle$~\cite{supp}. In brief, the stabilizer
generators $Z_1Z_2, Z_2Z_3, -Z_3Z_4, Z_4Z_5, Z_5Z_6$ lead to a
single encoded qubit
$$
  \alpha |(000)_u(111)_v\rangle + \beta |(111)_u(000)_v\rangle,
$$
supported by the two sites $u$ and $v$ jointly. Note the
antiferromagnetic ordering \cite{AKLT} among groups of three virtual
qubits. To reduce the support of this logical qubit to an individual
site, a measurement in the basis $\{|(000)_v\rangle \pm
|(111)_v\rangle\}$ is performed. The resulting state is $\alpha
|(000)_u\rangle \pm \beta |(111)_u\rangle$, with the sign ``$\pm$''
known from the measurement outcome. This is the proper encoding for
a domain consisting of a single site. Domains of more than two sites
are thereby reduced to  single sites.

To see that  $|\Psi({\cal{A}})\rangle$ is indeed equivalent under
local unitary transformations to an encoded graph state
$\overline{|G({\cal{A}})\rangle}$, we consider the example of four
domains $c$, $u$, $v$, $w$, each consisting of a single site of
${\cal{L}}$, where the POVM outcome is $z$ on the central domain $c$
and $x$ on all lateral domains $u$, $v$ and $w$; see
Fig.~\ref{merge}d. By similar arguments as above~\cite{supp}, the
operator ${\cal O}\equiv -X_1 X_{1'}X_2 X_{2'}X_3 X_{3'}$ is in the
stabilizer of $|\Psi({\cal{A}})\rangle$. Using the encoding in
Table~\ref{tbl:coding2}, i.e., with the encoded Pauli operators
$\overline{X}_c=Z_1Z_2Z_3$, $\overline{Z}_u=\pm X_{1'}$,
$\overline{Z}_v=\pm X_{2'}$, and $\overline{Z}_w=\pm X_{3'}$, we
find that ${\cal O}=\pm\overline{X}_c
\overline{Z}_u\overline{Z}_v\overline{Z}_w$ which is (up to a
possible sign) one  stabilizer generator defining the graph state;
see  Eq.~(\ref{eqn:stabilizerGen}).

 By the above construction, if two domains $u$,
$v$ are connected by an edge of multiplicity $m$, the inferred graph
state stabilizer generators will contain factors of
$\overline{X}_u{\overline{Z}_v}^m$ or
$\overline{X}_v{\overline{Z}_u}^m$. Rule R2 thus follows the
observation that $Z^2 = I$. Generalizing these  ideas, one can
rigorously prove that for any ${\cal{A}}$ of POVM outcomes, the
state $|\Psi({\cal{A}})\rangle$ is local-unitarily equivalent to an
encoded graph state $\overline{|G({\cal{A}})\rangle}$~\cite{arXiv}.
We shall denote by $|G({\cal A})\rangle$ the same graph state but
with  domains of single sites.

{\it Random graph states and percolation}. Whether or not typical graph states
$|G({\cal{A}})\rangle$ are universal resources hinges solely on connectivity
properties of $G({\cal{A}})$, and is thus a percolation problem~\cite{Perc}.
Specifically, for a large initial ${\cal{L}}$ the random graph state
$|G({\cal{A}})\rangle$ can, with close to unit probability, be efficiently
reduced to a large two-dimensional cluster state if the following properties
hold:
\begin{enumerate}
  \item[C1]{\label{C1}The distribution of the number of sites in a domain (i.e. domain size)
   is {\em{microscopic}}, i.e.,  the largest domain size can at most scale logarithmically with
     the total number of sites $|V({\cal{L}})|$ in the large ${\cal{L}}$ limit.}
  \item[C2]{\label{C2}The probability of the existence of a path through $G({\cal{A}})$ from the left to the right
  (or top to bottom) approaches unity in the limit of large ${\cal{L}}$.}
\end{enumerate}
Condition~C1 ensures that the graph $G({\cal{A}})$ remains
macroscopic if the original ${\cal{L}}$ was, and Condition~C2
ensures that the system is in the supercritical phase with a
macroscopic spanning cluster.

Together with planarity, which holds for all graphs $G({\cal{A}})$
by construction,  the conditions C1 and C2 are sufficient for the
reduction of the random graph state to a standard universal cluster
state. The proof~\cite{arXiv} extends a similar result already
established for site percolation on a square lattice~\cite{BPerc}.
The physical intuition comes from percolation theory. In the
supercritical phase (where there exists a macroscopic spanning
cluster and connects one boundary to the other), the spanning
cluster contains a subgraph which is topologically equivalent to a
coarse-grained two-dimensional lattice structure. This subgraph can
be carved out and subsequently cleaned off all imperfections by
local Pauli measurements, leading to a perfect two-dimensional
lattice.

\begin{figure}
\vspace{-0.5cm} {\includegraphics[width=8.5cm]{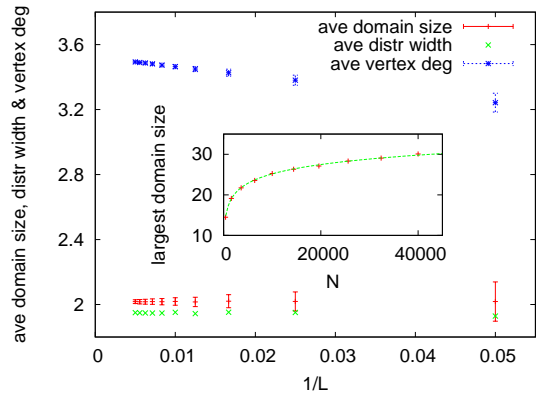}} \vspace{-1cm}
 \caption{(color online)  Average domain size, average width of domain size distribution,
 average degree of a vertex, and the largest domain size (inset) in the typical graphs
 vs. $L$, with $N=L^2$ being the total number
 of sites. For better discernibility,
 we suppress
 the errorbars for one set of data. The largest domain size scales with $N$
 as $3.337\ln(N)-5.566$.} \label{fig:avdeg}
\end{figure}

\begin{figure}
 {\includegraphics[width=8cm]{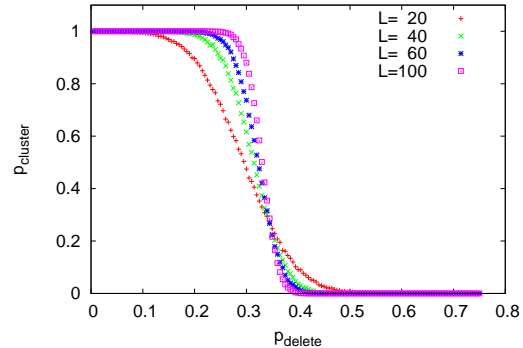}} \vspace{-1cm}
        \caption{(color online)  Percolation study of the random graphs of domains: probability of
        a spanning cluster $p_{\rm cluster}$ vs. that to delete a vertex  $p_{\rm delete}$.
        The threshold
        for destroying the spanning cluster is around $p_{\rm delete}\approx 0.33$ in deleting vertices.  } \label{fig:per}
\end{figure}
{\it Numerical results}. We used Monte Carlo simulations to sample
typical random graphs resulting from the POVM and compute their
properties. The simulations utilize a generalized Hoshen-Kopelman
algorithm~\cite{HoshenKopelman} to identify domains. {Due to the
entanglement in the AKLT state, the local POVM outcomes are
correlated which is fully taken into account in our simulations. In
particular, to sample typical POVM outcomes ${\cal A}$ correctly, we
use a Metropolis method to update configurations. For each site we
attempt to flip the type (either $x$, $y$ or $z$) to one of the
other two  equally and accept the flip with a probability $p_{\rm
acc}=\min\big\{1, 2^{|V'|-|{\cal E}'|-|V|+|{\cal E}|}\big\}$, where
$|V|$ and $|{\cal E}|$ denote the number of domains and inter domain
edges, respectively, before the flip and Rule 2, and similarly
$|V'|$ and $|{\cal E}'|$ for the flipped configuration~\cite{arXiv}.

We have analyzed lattices of size up to $200 \times 200$ sites. The
average degree of vertices in the typical random graphs is about
$3.52(1)$, when extrapolated to an infinite system size.
Furthermore, the typical random graphs retain large number of
vertices $|\bar{V}|= 0.495(2) N$, edges $|\bar{E}|=0.872(4) N$, and
independent cycles (or the Betti number) $\bar{B}= 0.377(2) N$,
where $N=L\times L$ is the total number of sites in the initial
honeycomb lattice. The size of the largest
 domain was never macroscopic and followed a logarithmic dependence on $N$.
 The average number
 of sites $v \in V({\cal{L}})$ contained in  a typical domain, when extrapolated to
 the infinite system, is $2.02(1)$ and the width in the domain size distribution is
 extrapolated to $1.94(1)$. Our simulations show that
 condition C1 holds; see Fig.~\ref{fig:avdeg}. For all  POVM outcomes sampled, a macroscopic cluster existed, allowing
 a horizontal and a
vertical traversing path through the resulting graphs $G({\cal{A}})$. This
shows that condition C2 holds.

{\em{Robustness.}} We now quantify how deep typical graphs
$G({\cal{A}})$ are in the supercritical phase of the percolation
transition. A first measure is the average vertex degree. A
heuristic argument based on a branching process suggests that a
graph has a macroscopic connected component whenever the average
vertex degree is $\bar{d}>2$~\cite{footnote}. In our case, the
typical graphs $G({\cal{A}})$ have an average degree of 3.52,
suggesting that the system is deep in the supercritical phase.
Furthermore, we randomly delete
 a fraction of vertices or edges from  $G({\cal{A}})$. On average,
 it requires a deletion probability as high as  $p_{\rm delete}^* =
0.33(1)$ for {vertices} (see Fig.~\ref{fig:per}) and $p_{\rm
delete}^* = 0.43(1)$ for {edges} (not shown) in order to the
spanning cluster. These numbers demonstrate the robustness of the
connectivity property.

{\it Concluding remarks}. We investigated MBQC on the AKLT states and
established one crucial missing ingredient in this area: the two-dimensional
spin-3/2 AKLT state on a honeycomb lattice is indeed a universal resource.
 The approach described in this work also applies to other trivalent lattices,
 such as the Archimedean lattices: $(3,12^2)$, $(4,6,12)$ and $(4,8^2)$,
 which have higher percolation thresholds than the honeycomb lattice.

 After the completion of our work, we learned of a similar result by Miyake
with a different approach~\cite{Miyake10}.

\smallskip \noindent {\bf Acknowledgment.} This work was supported by NSERC,
MITACS, CIFAR and the Sloan Foundation.

\appendix
\begin{widetext}
\section{ Projection onto the symmetric subspace of three
qubits}. The addition of angular momenta for
 three qubits (spin-1/2) gives rise to three subspaces:
 $\frac{1}{2}\otimes \frac{1}{2}\otimes\frac{1}{2}=\frac{1}{2}\oplus \frac{1}{2}\oplus\frac{3}{2}$.
The four basis states of the $S=3/2$ subspaces are $\ket{3/2,3/2}$,
$\ket{3/2,-3/2}$, $\ket{3/2,1/2}$ and $\ket{3/2,-1/2}$ (with the
quantization axis being assumed to be $z$), and they can be
expressed in terms of the three-qubit basis states (with
$\ket{0/1}\equiv \ket{\uparrow/\downarrow}=\ket{1/2,\pm1/2}$)
\begin{subequations}
\begin{eqnarray}
&&\ket{3/2,3/2}=\ket{000}, \ \ket{3/2,-3/2}=\ket{111},\\
&& \ket{3/2,1/2}=\ket{W}\equiv
\frac{1}{\sqrt{3}}(\ket{001}+\ket{010}+\ket{100}), \\
&&\ket{3/2,-1/2}=\ket{\overline{W}}\equiv\frac{1}{\sqrt{3}}(\ket{110}+\ket{101}+\ket{011}).
\end{eqnarray}
\end{subequations}
As can be clearly seen, states in this subspace are symmetric under
permutation of the three qubits and  the corresponding  projection
operator is
\begin{equation}
P_{S,v}\equiv|000\rangle\langle 000|+ |W\rangle \langle W|
+|\overline{W}\rangle \langle \overline{W}| + |111 \rangle\langle
111|.
\end{equation}

\smallskip {\it The POVM and the post-POVM state\/}.
 To show that the 2D AKLT state
of four-level spin-3/2 particles can be converted to a graph state
of two-level qubits, we need to preserve a local two-dimensional
structure at each site. This is achieved by a local generalized
measurement~\cite{NielsenChuang00}, also called
positive-operator-value measure (POVM), on every site $v$ on
${\cal{L}}$. The POVM consists of three rank-two elements
$F_{v,a}^\dagger F_{v,a}$ with $a=x, y,\,{\rm or}\, z$, and
\begin{subequations}
\label{POVM}
  \begin{eqnarray}
\!\!\!\!\!\!\!\!\!\!{F}_{v,z}&=&\sqrt{\frac{2}{3}}\big(\ketbra{000}+\ketbra{111}\big) \\
\!\!\!\!\!\!\!\!\!\!{F}_{v,x}&=&\sqrt{\frac{2}{3}}\big(\ketbra{+++}+\ketbra{---}\big)\\
\!\!\!\!\!\!\!\!\!\!{F}_{v,y}&=&\sqrt{\frac{2}{3}}\big(\ketbra{i,i,i}+\ketbra{-\!i,-\!i,-\!i}\big),
\end{eqnarray}
\end{subequations}
where $|0/1\rangle$, $\ket{\pm}\equiv(\ket{0}\pm\ket{1})/\sqrt{2}$
and $\ket{\pm i}\equiv (\ket{0}\pm i\ket{1})/\sqrt{2}$ are
eigenstates of the three Pauli operators $Z$, $X$ and $Y$,
respectively. Physically, $F_{v,a}$ is proportional to a projector
onto the two-dimensional subspace spanned by the $S_a=\pm 3/2$
states, i.e.,
\begin{equation}
{F}_{v,a}=\sqrt{\frac{2}{3}}\big(\ket{3/2,+3/2}_a\bra{3/2,+3/2}\,+\,\ket{3/2,-3/2}_a\bra{3/2,+3/2}\big),
\end{equation}
with $\hat{S}_{a}\ket{3/2,\pm3/2}_a= \pm 3/2 \ket{3/2,\pm3/2}_a$ and
$a$ indicating the quantization axis.

 The above POVM elements obey the
relation $\sum_{\nu \in \{x,y,z\}}F^\dagger_{v,\nu} F_{v,\nu} =
P_{S,v}=\openone_{S=3/2}$, i.e., project onto the symmetric
subspace, as required. The outcome $a_v$ of the POVM at site $v$ is
random, which can be $x$, $y$ or $z$. The resulting quantum state,
dependent on the random POVM outcomes ${\cal{A}}=\{a_v, v\in
V({\cal{L}})\}$,
\begin{equation}
  \label{CA0}
  |\Psi({\cal{A}})\rangle = \bigotimes_{v \in V({\cal{L}})} \!\! F_{v,a_v}\, |\Phi_{\rm AKLT}
\rangle= \bigotimes_{v \in V({\cal{L}})} \!\! F_{v,a_v}
\bigotimes_{e \in E({\cal{L}})} |\phi\rangle_e,
\end{equation}
where $|\phi\rangle_e$ denotes a singlet state
$(|01\rangle-|10\rangle)/\sqrt{2}$ on the edge $e$.

\smallskip{\it Example 1: qubit encoding\/}.
 Consider the case where two neighboring POVMs yield the same
outcome, say $z$; see Fig.~2c in the main text. Due to the
projections $F_{u,z}$ and $F_{v,z}$ (with $u=\{1,2,3\}$ and
$v=\{4,5,6\}$ each containing three virtual qubits), the operators
$Z_1Z_2$, $Z_2Z_3$, and $Z_4Z_5$, $Z_5 Z_6$ become stabilizer
generators of $|\Psi({\cal{A}})\rangle$, as, e.g.,
\begin{equation}
Z_1Z_2 F_{u,z}=F_{u,z} \end{equation}
 and hence
\begin{equation}
Z_1Z_2|\Psi({\cal{A}})\rangle=Z_1Z_2\bigotimes_{v \in V({\cal{L}})}
\!\! F_{v,a_v} \bigotimes_{e \in E({\cal{L}})}
|\phi\rangle_e=|\Psi({\cal{A}})\rangle.
\end{equation}
Moreover, the stabilizer $-Z_3Z_4$ of $|\phi\rangle_{34}$ (i.e.,
$-Z_3Z_4|\phi\rangle_{34}=|\phi\rangle_{34}$) commutes with
 $F_{u,z}\otimes F_{v,z}$, and thus remains a stabilizer element
for $|\Psi({\cal{A}})\rangle$,
\begin{equation}
-Z_3Z_4|\Psi({\cal{A}})\rangle=\bigotimes_{v \in V({\cal{L}})} \!\!
F_{v,a_v}(-Z_3Z_4) \bigotimes_{e \in E({\cal{L}})}
|\phi\rangle_e=|\Psi({\cal{A}})\rangle.
\end{equation}
 Therefore, regarding sites $u$ and $v$, $ |(000)_u(111)_v\rangle$ and $ |(111)_u(000)_v\rangle$ are the only
 two basis states that are stabilized (i.e., common eigenstates with eigenvalue equal to unity) by
 the above five operators, $Z_1Z_2, Z_2Z_3, -Z_3Z_4,
Z_4Z_5, Z_5Z_6$. This leads to a single encoded qubit
\begin{equation}
  \alpha |(000)_u(111)_v\rangle + \beta |(111)_u(000)_v\rangle,
\end{equation}
supported by the two sites $u$ and $v$ jointly. If more neighboring
sites share the same POVM outcome, the encoded qubit can be easily
extended.

\smallskip{\it Example 2: graph-state stabilizer generator\/}. To see
that $|\Psi({\cal{A}})\rangle$ is indeed equivalent under local
unitary transformations to an encoded graph state
$\overline{|G({\cal{A}})\rangle}$, we consider the example of four
domains $c$, $u$, $v$, $w$, each consisting of a single site of
${\cal{L}}$, where the POVM outcome is $z$ on the central domain $c$
and $x$ on all lateral domains $u$, $v$ and $w$; see Fig.~2d of the
main text. By direct computation we have that
\begin{equation}
[-X_1 X_{1'}, F_{u,x}]=[-X_2 X_{2'}, F_{v,x}]=[-X_3 X_{3'},
F_{w,x}]=0.
\end{equation}
Although $-X_{i} X_{i'}$ individually does not commute with
$F_{c,z}$,  the operator ${\cal O}\equiv X_1 X_{1'}X_2 X_{2'}X_3
X_{3'}$ does, as by direction computation,
\begin{equation}
X_1 X_2 X_3
(\ketbra{000}+\ketbra{111})=\ket{111}\bra{000}+\ket{000}\bra{111}=(\ketbra{000}+\ketbra{111})X_1
X_2 X_3,
\end{equation}
 we have
\begin{equation}
[X_1X_2X_3,F_{c,z}]=0.
\end{equation}
This shows that ${\cal O}$ is in the stabilizer of
$|\Psi({\cal{A}})\rangle$, i.e., ${\cal
O}|\Psi({\cal{A}})\rangle=|\Psi({\cal{A}})\rangle$. Using the
encoding in Table~1 of the main text, i.e., with the encoded Pauli
operators $\overline{X}_c=Z_1Z_2Z_3$, $\overline{Z}_u= X_{1'}$,
$\overline{Z}_v= X_{2'}$, and $\overline{Z}_w= X_{3'}$, we find that
${\cal O}=\overline{X}_c \overline{Z}_u\overline{Z}_v\overline{Z}_w$
which is (up to a possible sign) one stabilizer generator defining
the graph state.

\smallskip{\it The Hamiltonian and boundary conditions\/}. The
construction of the AKLT state by the valence-bond picture gives
rise to a state that is the ground state of the following
Hamiltonian
\begin{equation}
 H=\sum_{{\rm edge}\,\langle i,j\rangle}\hat{P}_{i,j}^{(S=3)}=
 \sum_{{\rm edge}\,\langle i,j\rangle}\Big[ \vec{S}_i\cdot \vec{S}_{j}+
 \frac{116}{243}(\vec{S}_i\cdot \vec{S}_{j})^2+\frac{16}{243}(\vec{S}_i\cdot \vec{S}_{j})^3+\frac{55}{108}
 \Big],
\end{equation}
where $\hat{P}_{i,j}^{(S=3)}$ is a projector of the neighboring
sites $i$ and $j$ onto a total $S=3$ subspace. In the case of the
periodic boundary condition, the AKLT is the unique ground state. In
the case of the open boundary condition, one can terminate every
boundary spin-3/2 by a spin-1/2, and add a corresponding term in the
Hamiltonain,
\begin{equation}
h_{i,i'}=P_{i,i'}^{(S=2)}=\frac{1}{2}\vec{S}_i\cdot\vec{s}_{i'}
+\frac{5}{4},
\end{equation}
where $\vec{S}_i$ is the spin-3/2 operator at the boundary and
$\vec{s}_{i'}$ is the associated spin-1/2 operator. These additional
Hamiltonian terms for all boundary pairs make the AKLT state a
unique ground state.

In the case of the open boundary condition, the planarity of the
graph is preserved even after the POVMs. However, in the case of the
periodic boundary condition, the underlying topology is that of a
torus. To make the graph of the corresponding graph state after the
POVMs be planar, one simply measures the logical $Z$ on sites along
the two independent cycles and this will cut the torus into a plane.

\end{widetext}


\begin{thebibliography}{99}
\bibitem{commentgap}
With an appropriate choice of boundary conditions, the AKLT state is
the unique ground state; see also Ref.~cite{supp}. In contrast to
1D, the 2D AKLT state on the honeycomb lattice was only shown to
possess exponentially decaying correlation functions~\cite{AKLT},
nevertheless suggesting
 a spectral gap.
\bibitem{footnote} This criterion, based on the critical degree $d_c$, is exact for random
graphs of uniform degree~\cite{Perc}. It holds surprisingly well for
lattice graphs, e.g., in honeycomb: $d_c=1.958$, Kagome:
$d_c=2.097$, square: $d_c=2$, triangular: $d_c=2.084$.
\end{thebibliography}
\end{document}